\PassOptionsToPackage{hyphens}{url}
\documentclass[a4paper, amsfonts, amssymb, amsmath, reprint, showkeys, nofootinbib, twoside, superscriptaddress]{revtex4-2}
\pdfoutput=1
\usepackage[english]{babel}
\usepackage[utf8]{inputenc}
\usepackage[colorinlistoftodos, color=green!40, prependcaption]{todonotes}
\usepackage[caption=false]{subfig}
\usepackage[pdftex, pdftitle={Article}, pdfauthor={Author}]{hyperref} 
\usepackage{amsthm}
\usepackage{mathtools}
\usepackage{physics}
\usepackage{xcolor}
\usepackage{graphicx}
\usepackage{subfig}
\usepackage[left=23mm,right=13mm,top=35mm,columnsep=15pt]{geometry} 
\usepackage{adjustbox}
\usepackage{placeins}
\usepackage[T1]{fontenc}
\usepackage{lipsum}
\usepackage{csquotes}
\usepackage{multirow}
\usepackage{braket}
\usepackage{comment}
\usepackage{bbold}

\begin{document}
\title{Requirements for upgrading trusted nodes to a repeater chain over 900 km of optical fiber}

\author{Francisco Ferreira da Silva}
    \thanks{These authors contributed equally.}
    \affiliation{QuTech, Delft University of Technology, Lorentzweg 1, 2628 CJ Delft, The Netherlands}
    \affiliation{Quantum Computer Science, EEMCS, Delft University of Technology, Lorentzweg 1, 2628 CJ Delft, The Netherlands}
    \affiliation{Kavli Institute of Nanoscience, Delft University of Technology, Lorentzweg 1, 2628 CJ Delft, The Netherlands}
    \affiliation{Corresponding author: \href{mailto:f.hortaferreiradasilva@tudelft.nl}{f.hortaferreiradasilva@tudelft.nl}}
\author{Guus Avis}
    \thanks{These authors contributed equally.}
    \affiliation{QuTech, Delft University of Technology, Lorentzweg 1, 2628 CJ Delft, The Netherlands}
    \affiliation{Quantum Computer Science, EEMCS, Delft University of Technology, Lorentzweg 1, 2628 CJ Delft, The Netherlands}
    \affiliation{Kavli Institute of Nanoscience, Delft University of Technology, Lorentzweg 1, 2628 CJ Delft, The Netherlands}
\author{Joshua A. Slater}
    \affiliation{QuTech, Delft University of Technology, Lorentzweg 1, 2628 CJ Delft, The Netherlands}
\author{Stephanie Wehner}
    \affiliation{QuTech, Delft University of Technology, Lorentzweg 1, 2628 CJ Delft, The Netherlands}
    \affiliation{Quantum Computer Science, EEMCS, Delft University of Technology, Lorentzweg 1, 2628 CJ Delft, The Netherlands}
    \affiliation{Kavli Institute of Nanoscience, Delft University of Technology, Lorentzweg 1, 2628 CJ Delft, The Netherlands}
    \affiliation{Corresponding author: \href{mailto:s.d.c.wehner@tudelft.nl}{s.d.c.wehner@tudelft.nl}}

\date{\today} 

\begin{abstract}
 We perform a numerical study of the distribution of entanglement on a real-world fiber grid connecting the German cities of Bonn and Berlin.
 The connection is realized using a chain of processing-node quantum repeaters spanning roughly 900 kilometers.
 We investigate how minimal hardware requirements depend on the target application, as well as on the number of repeaters in the chain.
 We find that requirements for blind quantum computing are markedly different than those for quantum key distribution, with the required coherence time being around two and a half times larger for the former.
 Further, we observe a trade-off regarding how target secret-key rates are achieved when using different numbers of repeaters:
 comparatively low-quality entangled states generated at a high rate are preferred for higher numbers of repeaters, whereas comparatively high-quality states generated at a lower rate are favored for lower numbers of repeaters.
 To obtain our results we employ an extensive simulation framework implemented using NetSquid, a discrete-event simulator for quantum networks.
 These are combined with an optimization methodology based on genetic algorithms to determine minimal hardware requirements.
 \end{abstract}

\maketitle

\section{Introduction}\label{sec:introduction}
The Quantum Internet promises to enable various novel applications that are provably impossible using the classical internet alone.
These include the secure distribution of secret keys~\cite{ekert1991quantum, bennett2020quantum}, distributed quantum computation~\cite{buhrman2003distributed} and secret sharing~\cite{hillery1999quantum}.
The requirements on hardware for a functioning quantum internet are application-dependent~\cite{wehner2018quantum}, but the generation and distribution of high-quality entanglement is necessary for all applications beyond quantum key distribution (QKD).

Recent years have seen major advances in quantum networking, with state-of-the-art experiments demonstrating entanglement generation at metropolitan distances~\cite{lago2021telecom} and a three-node quantum network in a lab~\cite{pompili2021realization}.
However, a large-scale quantum network requires long-distance entanglement generation, which has thus far not been realized.
This is due to the fact that the probability of not losing photons, the typical travelling carriers of quantum information, decays exponentially with the distance traversed in optical fiber.
Classically, signals are amplified to overcome loss. 
The same solution is impossible in the quantum case due to the no-cloning theorem~\cite{park1970, wootters1982, dieks1982}, which renders copying arbitrary quantum states impossible.

Quantum repeaters~\cite{briegel1998quantum, dur1999quantum} have been proposed as a possible solution for dealing with photon loss by introducing intermediate nodes between the two distant locations over which entanglement is to be established.
Multiple platforms have been considered for implementing quantum repeaters (see, e.g.,~\cite{munro2015inside,muralidharan2016optimal}), including ensemble-based systems~\cite{sangouard2011quantum, duanLongdistanceQuantumCommunication2001}, trapped ions~\cite{monroe2013, reiserer2015a, krutyanskiy2022telecom}, neutral atoms~\cite{reiserer2015a, langenfeld2021a} and color centers in diamond~\cite{ruf2021quantum}.
All but the first are examples of processing nodes.
These are capable of not only storing and transmitting information, but also of performing quantum gates.

Despite recent progress, a scalable quantum repeater has yet to be demonstrated.
Part of the challenge is that hardware requirements are not fully known.
There is a large body of work investigating such requirements (see~\cite{avis2022requirements} and references therein).
However, to the best of our knowledge, there has yet to be a study of how hardware requirements for chains of multiple repeaters, with their placement constrained by a real-world fiber grid, depend on (i) the number of repeaters used and (ii) the application for which the entanglement will be used.

It is likely that real-world deployment of quantum networks will make use of existing fiber infrastructure~\cite{rabbie2022}.
Previous work has shown that accounting for this fact significantly affects the hardware requirements for a single processing-node repeater setup~\cite{avis2022requirements}.
This emphasizes the importance of taking the constraints imposed by existing fiber grids into consideration when estimating repeater hardware requirements.
Moreover, existing fiber grids are already being used to deploy trusted-node networks throughout the world~\cite{stanley2022recent}.
These networks allow for the distribution of keys, albeit without end-to-end security~\cite{salvail2010security}.
A natural development in the deployment of quantum networks is to upgrade trusted-node networks by replacing them with quantum-repeater networks~\cite{wehner2018quantum}.
This provides further motivation for considering real-world fiber grids when studying hardware requirements.

Given that future quantum networks will be used for different applications, it is also pertinent to ask whether the required hardware quality depends on the application to be executed.
QKD is one of the best-known applications for quantum networks~\cite{ekert1991quantum, bennett2020quantum}.
In its entanglement-based version~\cite{bennett1992quantum}, two parties that share entangled pairs can perform measurements in two different bases and compare outcomes in order to distill a provably-secret key that can then be used for cryptographic protocols.
Verifiable blind quantum computation (VBQC) is a promising application that allows a client to execute a computation on a powerful remote server securely and privately~\cite{broadbent2009universal, leichtle2021verifying}.
QKD is the most often considered application when evaluating hardware requirements, but it is not clear whether its requirements are representative of those for other quantum-networking applications, particularly as QKD does not require multiple live entangled states at the same time (unlike, for example, VBQC).

In this work, we investigate minimal hardware requirements for connecting two end nodes which are separated by roughly 900 km of real-world optical fiber using a chain of processing-node quantum repeaters.
We study how these requirements change with the number of repeaters used and the imposed performance target.
In particular, we compare minimal hardware requirements for performing QKD at different rates (1, 10 and 100 Hz) and for performing VBQC.

\subsection{Setup}
\label{sec:setup}
\begin{figure}[!ht]
\centering
\includegraphics[width=\columnwidth]{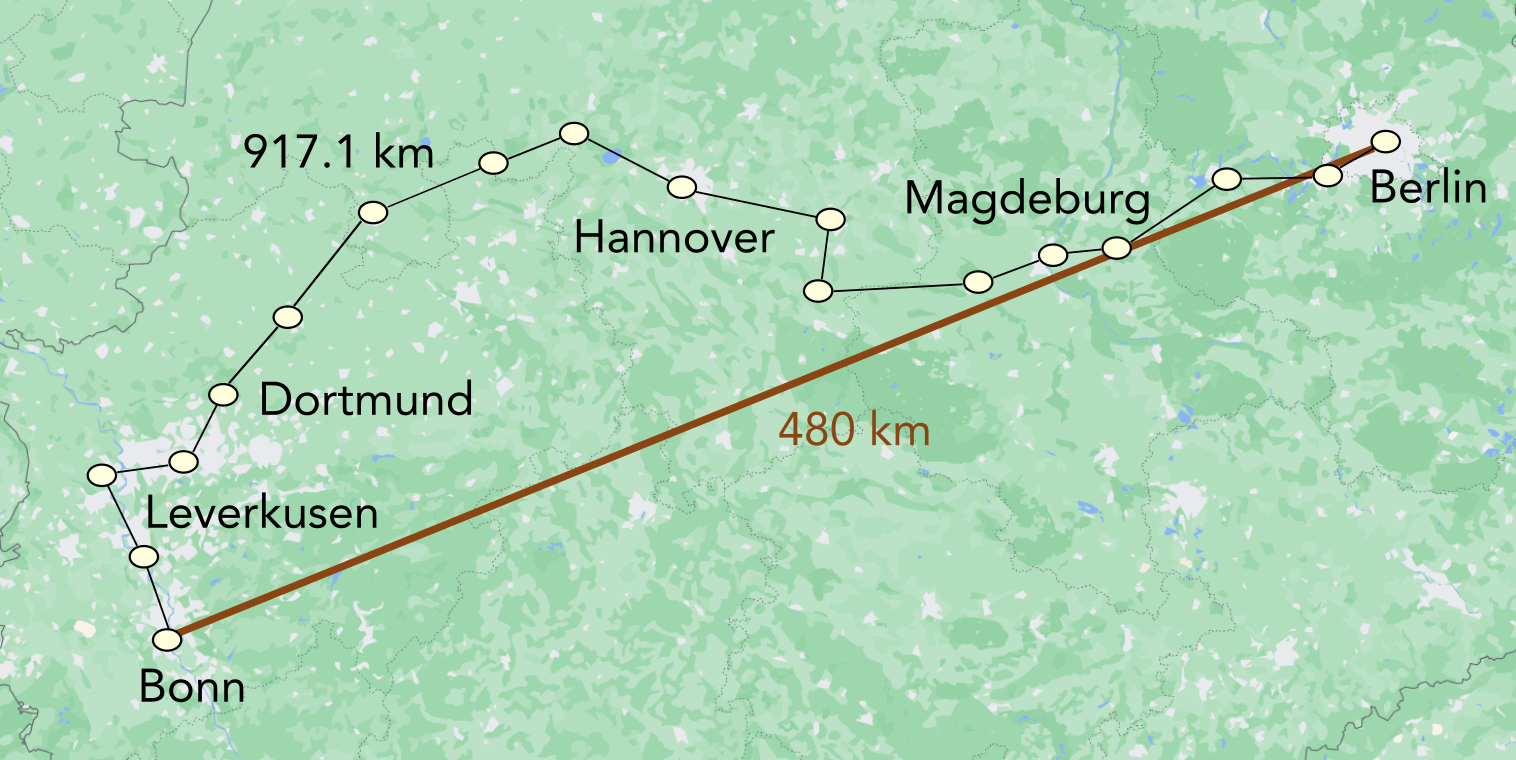}
\caption{Map of Germany overlaid with a depiction of the fiber path connecting the German cities of Bonn and Berlin that we investigate, provided by Deutsche Telekom (DT).
The white circles represent locations where DT plans to install trusted nodes and where, when building a repeater chain, processing nodes or heralding stations could be placed.
These locations are connected to each other through fiber drawn in black.
The maximum number of repeaters that can be placed between Bonn and Berlin in this fiber network is seven.
We consider all possible repeater placements, assuming that the heralding stations are placed as symmetrically as possible (there are 986 such placements).
The distance between Bonn and Berlin is 917.1 km via fiber, and approximately 480 km as the crow flies.
The reason for such a large difference between the two values is that other major German cities, such as Hannover and Dortmund, are connected through the fiber link as well.
}
\label{fig:bonn_berlin_satellite}
\end{figure}

We consider the quantum-network path depicted in Figure~\ref{fig:bonn_berlin_satellite}, with two end nodes situated in Bonn and Berlin separated by 917.1 km of optical fiber corresponding to 214.7 dB of attenuation (at a telecom wavelength of 1550 nm).
There are a total of sixteen locations between the end nodes where equipment can be placed, namely repeater nodes and heralding stations.
Throughout this paper we assume that such a heralding station must be placed between every pair or neighboring network nodes (i.e., end nodes or repeater nodes), as these are required when entanglement is generated between those nodes through the interference and measurement of entangled photons \cite{cabrillo1999creation, barrettEfficientHighfidelityQuantum2005, humphreys2018deterministic, pompili2021realization, bernien2013heralded, hensen2015loophole, stephensonHighRateHighFidelityEntanglement2020, krutyanskiy2023, avis2022requirements}.
This data has been provided to us by Deutsche Telekom (DT), Germany's largest telecommunications provider, which plans to install trusted nodes in the locations depicted in Figure~\ref{fig:bonn_berlin_satellite}.

We assume neighboring nodes perform heralded entanglement generation~\cite{azuma2022quantum, northup2014}.
That is, entanglement consists of a series of attempts, and at the end of each attempt the partaking nodes learn whether an entangled state was successfully created or not.
Examples of protocols for heralded entanglement generation are the double-click protocol~\cite{barrettEfficientHighfidelityQuantum2005, bernien2013heralded, hensen2015loophole, krutyanskiy2023, pfaff2014unconditional, stephensonHighRateHighFidelityEntanglement2020}, where photons are interfered and measured at a heralding station and success is declared in case two detectors click, the single-click protocol~\cite{cabrillo1999creation, humphreys2018deterministic, kalb2017entanglement, pompili2021realization, slodicka2013}, where photons are also interfered but success is only declared in case one detector clicks, and direct transmission of an entangled photon from one node to the next where it is stored in heralded quantum memory~\cite{bhaskar2020experimental, langenfeld2021, lin2009}.
Here, we employ a simplified model for heralded entanglement generation.
We do this so that the protocol and its interplay with other components of the repeater chain can be readily understood and our modelling is not overly platform specific.
First of all, we assume that each node can perform heralded entanglement generation with two neighbours in parallel, which is not currently possible for all quantum-repeater platforms~\cite{ruf2021quantum, avis2022requirements}.
Second, we model the elementary-link states $\rho$ that are created upon the completion of a successful attempt as depolarized Bell states, i.e.,
\begin{equation}
\rho =  W \ket{\phi^+}\bra{\phi^+} + \frac{1-W}{4}\mathbb{1},
\end{equation}
where $W$ is related to the fidelity $F$ to the ideal Bell state $\ket{\phi^+} = \frac{1}{\sqrt{2}}\left(\ket{00} + \ket{11}\right)$ as $F = (1 + 3W)/4$ and $\mathbb{1}$ is the four-dimensional identity matrix.
We note that real entangled states generated in quantum-repeater chains are often not depolarized states~\cite{hermans2023entangling, avis2022requirements}.
Yet, as depolarized Bell states represent a worst-case type of noise~\cite{horodeckiGeneralTeleportationChannel1999}, using a depolarizing model ensures that we will not find hardware requirements that are artificially low due to this simplification.
Third, we take the time $t_\text{attempt}$ required to perform one attempt to be given by
\begin{equation}
t_\text{attempt} = \frac L c ,
\end{equation}
where $L$ is the fiber distance between the two nodes and $c = 200,000$ km/s is the speed of light in fiber.
That is, it corresponds to the communication time associated with sending photons to a heralding station that is exactly in the center between two nodes and then receiving a message with the measurement outcome.
This is equivalent to the time required to directly transmit a photon from one node to the next.
In reality it may be longer, as the attempt time could be further limited by the rate of the photon source, local operations or the synchronization of emission times~\cite{pompili2021realization, pompiliExperimentalDemonstrationEntanglement2021}.
Finally, we take the success probability $p_\text{el}$ of each attempt to be
\begin{equation}
\label{eq:prob_el}
p_{\text{el}} = p_{\text{det}} \times 10^{-\frac{\alpha_{\text{att}}}{10}L}.
\end{equation}
Here, $p_\text{det}$ is the probability that an emitted photon that is led through fiber to a detector is detected, given that it is not lost while travelling in fiber.
This parameter combines multiple sources of loss, such as the detector's efficiency, the probability of emitting the photon in the right mode and the probability of successfully sending the photon into the fiber, but not the fiber's attenuation losses.
$\alpha_\text{att}$ is the fiber's attenuation coefficient (in dB/km).
Therefore, the success probability corresponds to the success probability of directly transmitting a photon between the nodes and measuring it there.
We note that for the double-click protocol the scaling with length would be the same, although the prefactor would be different ($p_\text{det}^2$ instead of $p_\text{det}$, as two photons must be detected).
For the single-click protocol the scaling would be more gentle (roughly replacing $L$ by $L/2$ in the exponential), and while the prefactor would be linear in $p_\text{det}$, there would also be a factor that depends on the device settings (specifically on the bright-state parameters chosen at both nodes, which tune a trade-off between success probability and state fidelity~\cite{pompili2021realization, childress2005, avis2022requirements}).
Additionally, we allow also for the possibility of multiplexed heralded entanglement generation~\cite{askarani2021entanglement, van2017multiplexed, sinclair2014spectral}.
This essentially consists of performing multiple attempts of generating the same elementary-link state in parallel.
Multiplexing can be done across multiple degrees of freedom, such as frequency, time or space.
We remain agnostic regarding how the multiplexing is performed, including it in our model as one parameter corresponding to the number of multiplexing modes used, $n$.
The probability of successfully generating an elementary link assuming the use of multiplexing is then the probability that at least one of the multiplexing modes succeeds:
\begin{equation}
p_{\text{multiple modes}} = 1 - \left(1 - p_{\text{el}}\right)^n.
\end{equation}

The nodes implement a swap-asap protocol~\cite{coopmans2021netsquid, inesta2022}.
That is, as soon as a node holds two entangled states, one shared with each of its neighbours, it performs an entanglement swap in order to create an entangled state spanning a larger distance.
We assume this swap is realized deterministically, since we are modelling processing nodes that can implement a swap using quantum gates and measurements on their processors.
It may however introduce noise, which we model as depolarizing.
We quantify how well the swap can be performed using the swap-quality parameter $s_q$.
The $d$-dimensional depolarizing noise channel of parameter $p$ acts on a state $\rho$ as follows,
\begin{equation}
\label{eq:depolarizing}
\rho \rightarrow p \rho + (1-p) \frac{\mathbb{1}}{d}.
\end{equation}
This means that, with probability $p$, $\rho$ is left unchanged, and with probability $1-p$ it is mapped to the maximally-mixed state, i.e., all information is lost.
Then, we model entanglement swapping as a two-qubit depolarizing channel (i.e., $d = 4$) with parameter $p = s_q$ followed by a perfect entanglement-swapping operation (i.e., a measurement in the Bell basis~\cite{bennett1993teleporting}).
We assume that the gates and measurements applied by the end nodes when executing QKD and VBQC are noiseless and instantaneous.
States stored in memory undergo decoherence, which we model as exponential depolarizing noise, i.e.,
\begin{equation} \label{eq:memory_decoherence}
    \rho \rightarrow e^{-t/T} \rho + \left(1 - e^{-t/T}\right)\frac{\mathbb{1}}{d},
\end{equation}
where $t$ is the time for which the state $\rho$ has been held in memory and $T$ is the memory's coherence time.
To combat the effects of memory decoherence, entangled states are discarded after a local cut-off time.
The cut-off time is defined as follows:
a timer starts once a state is created in memory through the successful generation of an elementary link.
If the timer reaches the local cut-off time, the state is discarded.
That is, the qubit holding the state is reset.
Additionally, the node sends a classical message along the chain so that the qubit with which the first qubit was entangled can also be reset.
As a result, a number of elementary links in the chain must be regenerated (with the exact number depending on how far away the entangled qubit was).

\subsection{Applications}
Having discussed our modelling of the entanglement generation process between Bonn and Berlin, we turn to the applications that will make use of the entanglement, QKD and VBQC.
We investigate the BB84 QKD protocol~\cite{bennett2020quantum} (in its entanglement-based form~\cite{bennett1992quantum}) between the end nodes situated in Bonn and Berlin.
We record the entanglement generation rate and estimate the quantum bit error rate (QBER) that would have been obtained when measuring the generated state in order to estimate the achievable asymptotic secret-key rate (SKR) as per the following equation~\cite{shor2000simple}:
\begin{equation}
\label{eq:skr}
\text{SKR} = E_R \cdot \max\Big(0, \big(1 - 2 H(Q)\big)\Big),
\end{equation}
where $E_R$ is the entanglement-generation rate, $H(p) = -p \log_2(p) - (1-p) \log_2(1-p)$ is the binary entropy function and $Q$ is the QBER.
We note that all the noise sources we consider are depolarizing, hence the entangled states generated will be of the form of the state shown in Equation~\ref{eq:depolarizing}.
Therefore, the QBER is the same irrespective of the measurement basis.
The end nodes do not wait until end-to-end entanglement is established before measuring their qubits.
Instead, they measure them as soon as they have established entanglement with their nearest neighbours, as this minimizes the amount of time states spend in memory, resulting in laxer hardware requirements.

We also investigate a two-qubit version of the VBQC protocol introduced in~\cite{leichtle2021verifying}.
In such protocols, a client wishes to delegate a computation to a powerful remote server in a secure and verifiable fashion~\cite{broadbent2009universal}.
In particular, we investigate the repeated execution of test rounds of the protocol, which consist of the server performing a controlled-Z gate followed by a measurement. 
In these rounds the client knows the computation's expected outcome, and can therefore compare them to the observed outcomes.
Under the assumption of an honest server, wrong outcomes are a result of noise.
We call this the \textit{BQC test protocol}.
The fraction of successful BQC test protocol rounds is therefore a metric for the quality of the entanglement used for transmitting qubits.
We define the success rate as the number of rounds of the protocol that can be executed with a successful result per time unit.
More concretely, if $p_s$ is the success probability of a test round and $R_{rounds}$ is the rate at which rounds can be executed, the BQC-test-protocol success rate is given by:
\begin{equation} \label{eq:BQC_success_rate}
    R_{\text{BQC}} = R_{rounds} \cdot p_s.
\end{equation}
While the BQC test protocol is in and of itself not an interesting application of a quantum network, it can be considered a benchmark for how well the network is suited to VBQC and possibly other multi-qubit applications.
The fact that, in contrast with QKD, it requires the distribution of multiple entangled states and the storage of qubits between rounds makes it a more meaningful benchmark for quantum-network applications that require multiple live qubits contemporaneously.
Further details on the BQC test protocol can be found in Appendix \ref{app:bqc_test_protocol}.

The two applications we have just introduced define our performance targets.
\subsection{Minimal hardware requirements}
\label{sec:defining_minimal_hardware_requirements}
We wish to find the \textit{minimal hardware requirements} that are needed to realize different target SKRs and BQC-test-protocol success rates.
These correspond to the minimal improvements over state-of-the-art hardware parameters that enable meeting the targets.
We phrase the problem of finding minimal hardware requirements as a constrained optimization problem.
Namely, we wish to minimize the hardware improvement while ensuring that the constraint induced by the performance target is met.
This constraint is relaxed through a process known as scalarization~\cite{pascoletti1984scalarizing, schaffer1985some}, resulting in a single-objective optimization problem, in which the quantity to be minimized is the sum of the cost associated to the hardware improvement and a penalty term for the rate target.
The resulting cost function is given by:
\begin{equation} \label{eq:cost_function}
\begin{aligned}
    C &= w_1 \big(1 + \left(R_{target} - R_{real}\right) \big)^2\\
    &\cdot \Theta \left(R_{target} - R_{real}\right) \\
    &+ w_2 H_C\left(x_1, ..., x_N\right),
\end{aligned}
\end{equation}
where $H_C$ is the hardware improvement cost associated to parameter set $\{x_1, ..., x_N\}$, $w_i$ are the weights assigned to the objectives, $\Theta$ is the Heaviside step function, $R_{target}$ is the rate target and $R_{real}$ is the rate of application execution achieved by the parameter set.
We note that $R_{real}$ and $R_{target}$ can be either a SKR or a BQC-test-protocol success rate.
$H_C$ maps sets of hardware parameters to a number, the cost, which represents how large of an improvement over the state-of-the-art they represent.
In order to compute this cost consistently across different parameters, we use no-imperfection probabilities as done in~\cite{avis2022requirements}.
By no-imperfection probability, we mean the probability that there is no error or loss associated to a given parameter.
For example, the no-error probability associated to a photon detection probability $p_{\text{det}}$ (defined in Section~\ref{sec:setup}) of $0.1$ is $0.1$.
For the no-error probabilities associated to the other hardware parameters, see Appendix \ref{app:no_imperfection_probs}.
We say that a parameter is improved by a factor of $k$ if its no-imperfection probability becomes $\sqrt[k]{p_{ni}}$, with $p_{ni}$ being the state-of-the-art no-imperfection probability.
For example, improving the no-imperfection probability of $0.1$ associated to $p_{\text{det}} = 0.1$ by a factor of $5$, we get a no-imperfection probability of $\approx 0.63$, corresponding to $p_{\text{det}} \approx 0.63$.
The hardware cost associated to a set of parameters is given by the sum of the improvement factors of the parameters.
The weights $w_i$ are chosen such that the term of the overall cost function corresponding to meeting the rate target is always larger than the one corresponding to the hardware cost, ensuring that even though we have relaxed the constraints by scalarizing, we are still effectively requiring that the minimal hardware requirements are such that the performance target is met.
To ensure this, we picked $w_1$, $w_2 \gg w_3$, such that $w_1 \big(1 + \left(R_{target} - R_{real}\right)^2 \big) \Theta \left(R_{target} - R_{real}\right)\gg w_2 H_C\left(x_1, ..., x_N\right)$.
Specifically, we set $w_1 = 1 \times 10^{100}$ and $w_2 = 1$.
No particular heuristic was used to select these numbers.

We note that the hardware cost is meant only to represent a measure of the hardness of improving the hardware to a certain level, and not any form of monetary cost.
At present quantum repeater systems are research setups, with commercial solutions only starting to emerge.
Therefore, assigning any specific commercial cost numbers would be too speculative at this point, and would require an in-depth study outside the scope of this project.

\subsection{State-of-the-art parameters}

Computing minimal hardware requirements as described in Section~\ref{sec:defining_minimal_hardware_requirements} is done with respect to a baseline over which we are improving.
In this work, this baseline consists of parameters measured for color centers in diamond, as they are physical systems using which various quantum-networking primitives have been demonstrated.
These include long-lived quantum memories~\cite{bradley2022robust}, remote entanglement generation~\cite{bernien2013heralded,hensen2015loophole}, quantum teleportation~\cite{pfaff2014unconditional}, entanglement distillation~\cite{kalb2017entanglement}, entanglement swapping~\cite{hermans2022qubit} and a three-node network~\cite{pompili2021realization}.
We do not impose that all parameters must have been demonstrated in the same experiment or even with the same color center.
The parameters we consider are shown in Table~\ref{tab:baseline_parameters}.
\begin{table}[!ht]
\begin{tabular}{|c|c|}
\hline
Parameter                    & Value                   \\ \hline
Coherence time               & 1 s~\cite{bradley2019ten}                     \\ \hline
Number of multiplexing modes & 1                       \\ \hline
Fidelity of elementary links & 0.83~\cite{hensen2015loophole}                    \\ \hline
Photon detection probability & 0.255~\cite{bhaskar2020experimental} \\ \hline
Swap quality                 & 0.83~\cite{kalb2017entanglement, taminiau2014universal}                    \\ \hline
\end{tabular}
\caption{State-of-the-art color-center parameters.
We note that not all of these parameter values have been realized in a single experiment.
We have number of modes as $1$ without reference because to the best of our knowledge multiplexed entanglement generation has not been demonstrated using color centers.}
\label{tab:baseline_parameters}
\end{table}
Details on how these parameters were determined can be found in Appendix~\ref{sec:appendix_baseline_parameters}.

\subsection{Determining minimal hardware requirements}
In order to determine minimal hardware requirements, we need to (i) be able to evaluate how a given set of hardware parameters performs and (ii) optimize over the parameter space to find the parameters that minimize the requirements while still performing adequately (i.e., the parameters that minimize the cost function defined in Equation \eqref{eq:cost_function}).

We evaluate the performance of hardware parameters using general processing-node repeater-chain simulations developed in NetSquid, a discrete-event based quantum-network simulator~\cite{coopmans2021netsquid}.
The simulations are general in the sense that they can be used to investigate swap-asap repeater chains of arbitrary size and spacing (i.e., nodes and heralding stations need not be equidistant).
They take into account time-dependent noise, classical control communication and the constraints imposed by a real-world fiber network.
The code for executing such simulations has been made publicly available at~\cite{netsquid-qrepchain}
and is largely based on the simulations first introduced in~\cite{avis2022requirements}.
Our code that utilizes these simulations to produce the results here presented can be found at~\cite{deutsche_telekom_code} (and the corresponding data at~\cite{2023requirementsreplicationdata}).

Given that we can evaluate the performance of any parameter set on the Bonn-Berlin path, we perform parameter optimization using a genetic algorithm~\cite{da2021optimizing} to minimize the cost function defined in Section~\ref{sec:defining_minimal_hardware_requirements} using the high-performance computing cluster Snellius.
For further details, see Appendix~\ref{sec:optimization_method}.

\section{Impact of number of repeaters on hardware requirements}\label{sec:results_chain_parameters}
In this section we answer the question of how hardware requirements are affected by the number of repeaters deployed in a quantum network.
Specifically, we investigate the minimal hardware requirements for performing BB84 between the German cities of Bonn and Berlin at a key rate of 10 Hz.
We assume the cities are connected by the network path shown in Figure~\ref{fig:bonn_berlin_satellite}.
We determine what these minimal requirements are in two cases:
(i) optimizing over the number of repeaters and
(ii) restricting the number of repeaters to specific values.
In both cases we optimize over the placement of the repeaters.

\subsection{Absolute minimal number of multiplexing modes}

Before determining minimal requirements, we aim to answer the question of what are the absolute minimal number of multiplexing modes required to perform QKD between the German cities of Bonn and Berlin at rates of 1, 10 and 100 Hz.
By \textit{absolute minimal} number of multiplexing modes, we mean the minimum number of multiplexing modes that is required if the only source of imperfection in the setup is fiber attenuation.
This provides a lower bound on the number of multiplexing modes in the minimal hardware requirements, as the introduction of other hardware imperfections can only lead to more stringent demands on the number of modes.
We emphasize that for the purposes of answering this question we are temporarily setting aside the real-world fiber path introduced in Figure~\ref{fig:bonn_berlin_satellite}.
Instead, we are going to consider a \textit{symmetrized} version of that path.
By this we mean a path with the same total length and attenuation, but in which nodes and heralding stations are placed equidistantly, and where the attenuation is evenly distributed throughout the path, i.e., all elementary links have the same attenuation.
The reason for doing so is that the minimal number of modes for this path is a lower bound for the same quantity on any other path with the same total length and attenuation.
To see this, we note that it has been shown in~\cite{avis2023asymmetry} that repeater chains of the type studied here perform best when all nodes are positioned as symmetrically as possible.
This implies that such a chain will have less stringent hardware requirements to attain a given performance target in comparison to chains which are subject to real-world restrictions such as the ones imposed by the fiber path shown in Figure~\ref{fig:bonn_berlin_satellite}, and, therefore, also less stringent requirements on the number of multiplexing modes.

Determining the absolute minimal number of multiplexing modes serves two purposes.
First, it allows us to limit the search space of the optimization we run for finding minimal hardware requirements.
Second, it gives us a general idea of how many repeaters might be required to achieve the target with reasonable hardware demands.
For example, if for a specific number of repeaters hundreds of thousands of multiplexing modes are required to meet the target without any noise sources, that may indicate that using that number of repeaters is not practically feasible.

\begin{figure}[!htpb]
\centering
\includegraphics[width=\columnwidth]{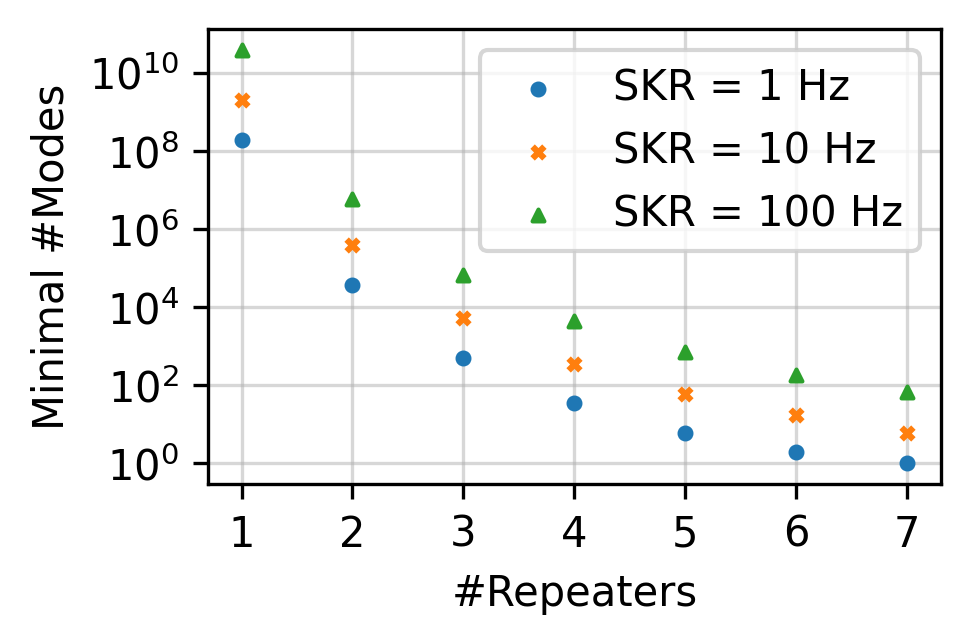}
\caption{Minimal number of multiplexing modes required to achieve 1, 10 and 100 Hz of SKR over 917.1 km of fiber with a total of 214.7 dB of attenuation, corresponding to a symmetrized version of the path between Bonn and Berlin that we investigate.
That is, for $N$ repeaters, the symmetrized path has $N+1$ elementary links, each of length $917.1/(N+1)$ km and of attenuation $214.7/(N+1)$ dB.
We assume that there are no hardware imperfections, and that repeaters are uniformly spaced.
}
\label{fig:absolute_minimal_number_of_modes}
\end{figure}

In Figure~\ref{fig:absolute_minimal_number_of_modes} we show the absolute minimal number of modes required to distribute secret key at rates of 1, 10 and 100 Hz using BB84 in the symmetrized Bonn - Berlin path.
We find that more multiplexing modes are required for higher rate targets, and that this number grows superexponentially as the number of repeaters decreases, so as to counteract the effects of photon loss in fiber.
Further, we find that achieving a SKR of 10 Hz with fewer than 3 repeaters requires hundreds of thousands of multiplexing modes even in the absence of any sources of noise.
As the hardware cost (defined in Section~\ref{sec:defining_minimal_hardware_requirements}) associated with so many multiplexing modes far outweighs typical values for the minimal total hardware cost found for three or more repeaters we limit the rest of our investigation to configurations with three or more repeaters.

\subsection{Minimal hardware requirements for quantum-key distribution}
\label{sec:min_requirements_qkd}
We now turn our attention to the minimal hardware requirements for performing quantum-key distribution at a rate of 10 Hz using the BB84 protocol.
In particular, we investigate them along the path connecting Bonn and Berlin depicted in Figure~\ref{fig:bonn_berlin_satellite}.
As Figure~\ref{fig:absolute_minimal_number_of_modes} illustrates, the number of repeaters used can have a considerable impact on the hardware requirements.
Further, it is expected that the same is true for the placement of repeaters and heralding stations~\cite{avis2022requirements, avis2023asymmetry}.
With this in mind, we ask two questions:
(i) what are the minimal hardware requirements when allowing for the placement of up to the largest number of repeaters that fits in the fiber path (seven) and
(ii) what are the minimal hardware requirements when restricting the maximum number of repeaters to five.
We expect that this will lead to different parameter regimes, illustrating two possible directions towards achieving the target performance.

\begin{figure}[!htpb]
\centering
\includegraphics[width=\columnwidth]{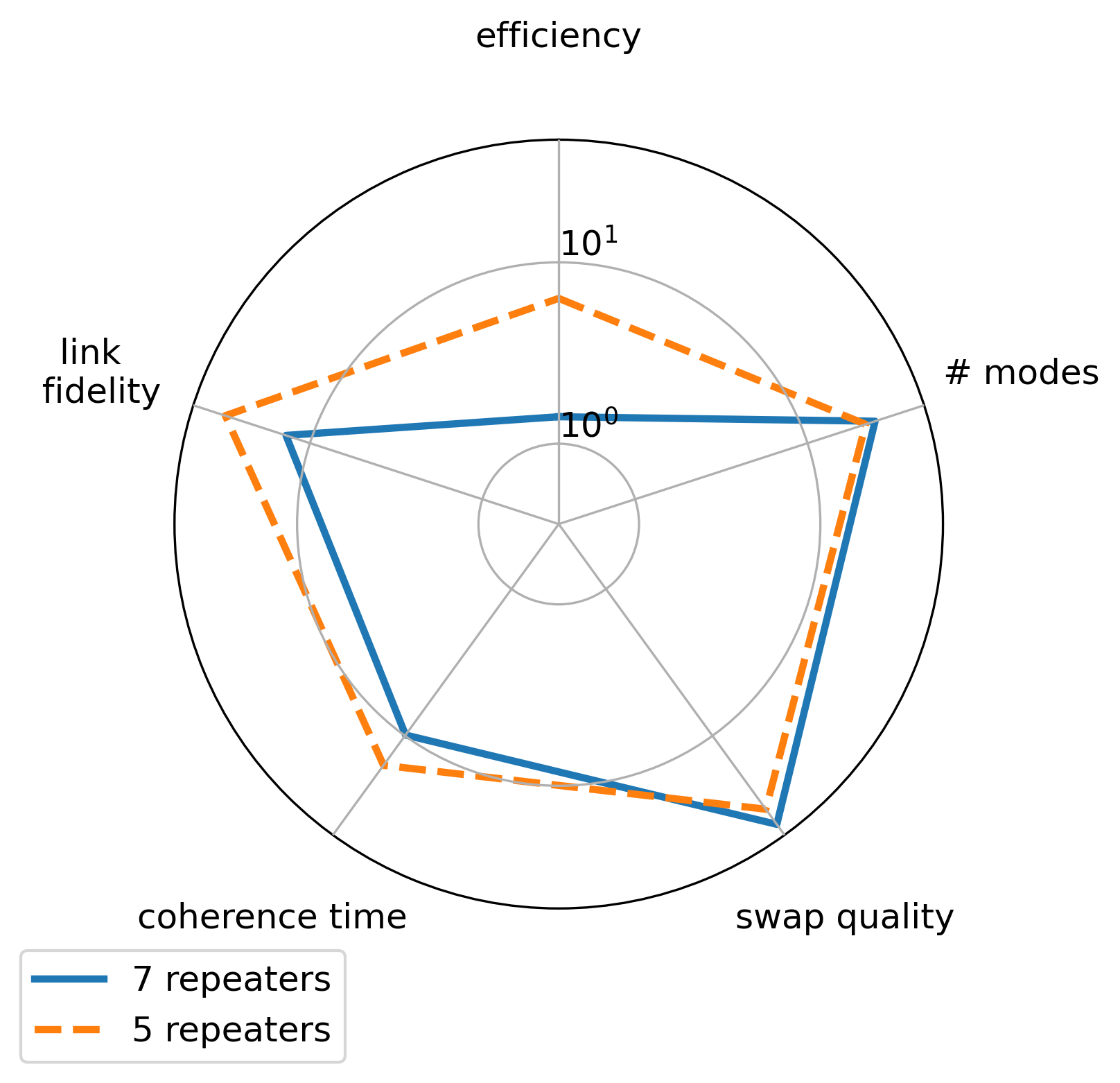}
\caption{Directions along which hardware must be improved to enable attaining a secret-key rate of 10 Hz between the German cities of Bonn and Berlin.
The blue (orange) line was obtained by performing an optimization in which the algorithm was allowed to use a maximum of seven (five) repeaters.
The further away the line is from the center of the plot towards a given parameter, the more that parameter must be improved with respect to the current state-of-the-art.
Improvement is depicted for the following parameters, clockwise from the top: overall photon detection probability excluding attenuation in fiber, number of multiplexing modes, fidelity of entanglement swap, coherence time of memory qubits and fidelity of elementary links.
Note the use of a logarithmic scale.
}
\label{fig:minimal_requirements_free_some_repeaters}
\end{figure}

In Figure~\ref{fig:minimal_requirements_free_some_repeaters} we show the directions along which hardware must be improved for distributing secret key at rates of 10 Hz using BB84 in the network path connecting Bonn and Berlin.
The corresponding minimal hardware requirements can be found in Table~\ref{tab:minimal_requirements_values}.
\begin{table*}[!ht]
\begin{tabular}{|c|cccc|c|}
\hline
Application                  & \multicolumn{4}{c|}{QKD}                                                                     & BQC   \\ \hline
Rate (Hz)                    & \multicolumn{1}{c|}{1}     & \multicolumn{2}{c|}{10}                                 & 100   & 10    \\ \hline
Number of repeaters          & \multicolumn{1}{c|}{7}     & \multicolumn{1}{c|}{Max 5} & \multicolumn{1}{c|}{Max 7} & 7     & 7     \\ \hline
Coherence time (s)           & \multicolumn{1}{c|}{1.81}  & \multicolumn{1}{c|}{4.23}  & \multicolumn{1}{c|}{3.14}  & 10.1  & 7.99  \\ \hline
Number of multiplexing modes & \multicolumn{1}{c|}{175}   & \multicolumn{1}{c|}{544}   & \multicolumn{1}{c|}{592}   & 799   & 172   \\ \hline
Fidelity of elementary links & \multicolumn{1}{c|}{0.989} & \multicolumn{1}{c|}{0.995} & \multicolumn{1}{c|}{0.987} & 0.996 & 0.845 \\ \hline
Photon detection probability $p_{\text{det}}$ & \multicolumn{1}{c|}{0.604} & \multicolumn{1}{c|}{0.785} & \multicolumn{1}{c|}{0.360} & 0.804 & 0.552 \\ \hline
Swap quality                 & \multicolumn{1}{c|}{0.996} & \multicolumn{1}{c|}{0.996} & \multicolumn{1}{c|}{0.997} & 0.997 & 0.881 \\ \hline
\end{tabular}
\caption{Minimal hardware requirements to achieve 1, 10 and 100 Hz of secret-key rate and 10 Hz of blind quantum computing test protocol success rate between the German cities of Bonn and Berlin.
The photon detection probability $p_{\text{det}}$ is the probability of a photon being detected given that it is not lost in fiber.
It combines multiple sources of loss, such as the detector's efficiency, the probability of emitting the photon in the right mode and the probability of successfully sending the photon into the fiber.
More details can be found in Section~\ref{sec:setup} and Appendix~\ref{sec:appendix_baseline_parameters}.}
\label{tab:minimal_requirements_values}
\end{table*}
In each case we find that the hardware requirements are minimized when the number of repeaters used is maximized.
That is, for seven repeaters in case (i) and five repeaters in case (ii).
Hardware requirements are more stringent in case fewer repeaters are used.
In particular, the overall photon detection probability excluding attenuation in fiber must be improved to a much larger degree (0.79 vs 0.36) if only five repeaters are used.
This is needed to overcome the increased attenuation losses associated with the longer elementary links.
The coherence time required when using five repeaters is also larger than the time required when using seven repeaters (4.2 s vs 3.1 s).
This can be explained by the fact that keeping the entanglement-generation rate high is more costly in case of five repeaters.
Therefore keeping the QBER small to extract as many secret bits as possible from each entangled state is more valuable.
Furthermore, since the entanglement-generation rate is smaller for five repeaters, qubits are stored for longer times before they can be swapped and hence a larger coherence time is required to achieve the same QBER.
We study this interplay further in Section~\ref{sec:qber_rate_tradeoff}.
Finally, we notice that while the requirements on most hardware parameters are more stringent for five repeaters as compared to seven repeaters, this is not the case for the requirement on the swap quality.
In fact, the requirement on the swap quality is even slightly looser for five repeaters (0.996 vs 0.997).
This is explained by the fact that when there are more repeaters, there are more entanglement swaps associated with every end-to-end entangled state.
Therefore, when there are more repeaters the final error rate is more sensitive to noise in the swaps, creating a larger incentive to improve the associated parameter in the seven-repeater case as compared to the five-repeater case.

\subsection{Secret-key rate: quantum-bit error rate and entanglement generation rate}
\label{sec:qber_rate_tradeoff} 
\begin{figure}[!ht]
\includegraphics[width=\columnwidth]{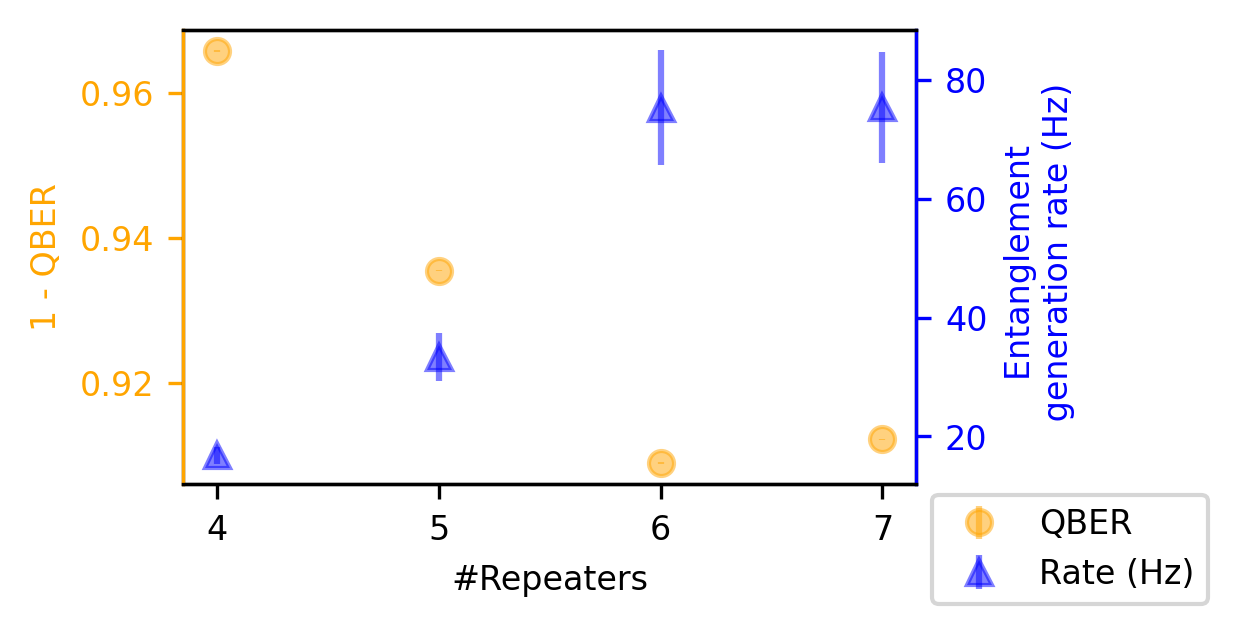}
\caption{QBER and entanglement generation rate obtained with the minimal hardware requirements to achieve 10 Hz of SKR in the Bonn - Berlin setup with different numbers of repeaters, up to seven, the maximum allowed in the setup we study.
The error bars are given by the standard error of the mean.
Each data point corresponds to 2000 simulations of an entanglement-based BB84 protocol.
}
\label{fig:qber_rate_repeaters}
\end{figure}

A specific value for the SKR can be obtained through many different pairs of values for the entanglement-generation rate and the QBER, as follows from Equation \eqref{eq:skr}.
This opens up a trade-off between the entanglement generation rate and the QBER, as briefly discussed in Section~\ref{sec:min_requirements_qkd}.
Here, we investigate this trade-off more deeply by repeating our process for determining minimal hardware requirements to achieve an SKR of 10 Hz while keeping the number of repeaters a fixed parameter.
We did this for 4, 5, 6 and 7 repeaters.
For each case, we still optimize over all possible placements of the repeaters in the fiber grid.
In Figure~\ref{fig:qber_rate_repeaters} we show the QBER and entanglement-generation rate achieved with the minimal hardware requirements for the best setup found by our optimization procedure for varying number of repeaters.
We observe two different regimes.
For 4 and 5 repeaters, which we name the `few-repeater' regime, we find a low QBER ($\sim5\%$) and an entanglement-generation rate of 20 - 30 Hz.
On the other hand, for 6 and 7 repeaters, i.e., the `many-repeater' regime, we find a comparatively higher QBER ($\sim9\%$) and an entanglement generation rate of almost 80 Hz.
In other words, in the many-repeater regime, distributing many entangled pairs of comparatively lower quality requires less hardware improvement. 
On the other hand, in the few-repeater regime it seems to be more feasible to distribute fewer pairs of comparatively higher quality.
As the number of repeaters used decreases, it becomes harder to overcome the effect of fiber attenuation, which makes improving the quality of the entangled states delivered a more attractive option for increasing the SKR.

We finalize by remarking that, perhaps surprisingly, the variance in the time it takes to distribute one entangled state appears to grow as the number of repeaters in the chain increases (as shown by the increasing error bar on the rate in Figure \ref{fig:qber_rate_repeaters}).
While interesting, further investigation is beyond the scope of this work.

We show the repeater placement corresponding to the minimal hardware requirements found when optimizing over the number of repeaters and their placement in Appendix~\ref{sec:appendix_best_rep_placement}.

\section{Impact of target on hardware requirements}\label{sec:results_target}
We now turn our attention to the impact of the performance target on the hardware requirements.
We approach this from two angles: (i) the impact of varying the SKR target and (ii) the impact of holding the required rate constant but changing the target quantum-network application.
It is clear that, given the same repeater chain, increasing the target rate will lead to more stringent requirements.
However, it is not \textit{a priori} obvious if the relative importance of the different parameters will change as the target rate is increased.
It is further not obvious how changing the target application impacts the hardware requirements.
These are questions of practical relevance:
given that one wishes to build a repeater chain capable of distributing entanglement to perform QKD at a rate of 100 Hz, it seems crucial to know whether building a repeater chain for performing QKD at a rate of 1 Hz is a step in the right direction.
In other words, this investigation can shed light on whether the process of improving hardware for quantum-repeater chains should be approached incrementally.
The same question holds for the different target applications.
It is likely that quantum repeaters will initially be used for QKD as they begin to replace their trusted-node predecessors, and only progressively start to be used for applications that require multiple live qubits.
We would then like to know whether the hardware improvements necessary to perform QKD using quantum repeaters are similar to the ones for multi-qubit applications.

\subsection{Requirements for different secret-key-rate targets}
\begin{figure}[!ht]
\includegraphics[width=\columnwidth]{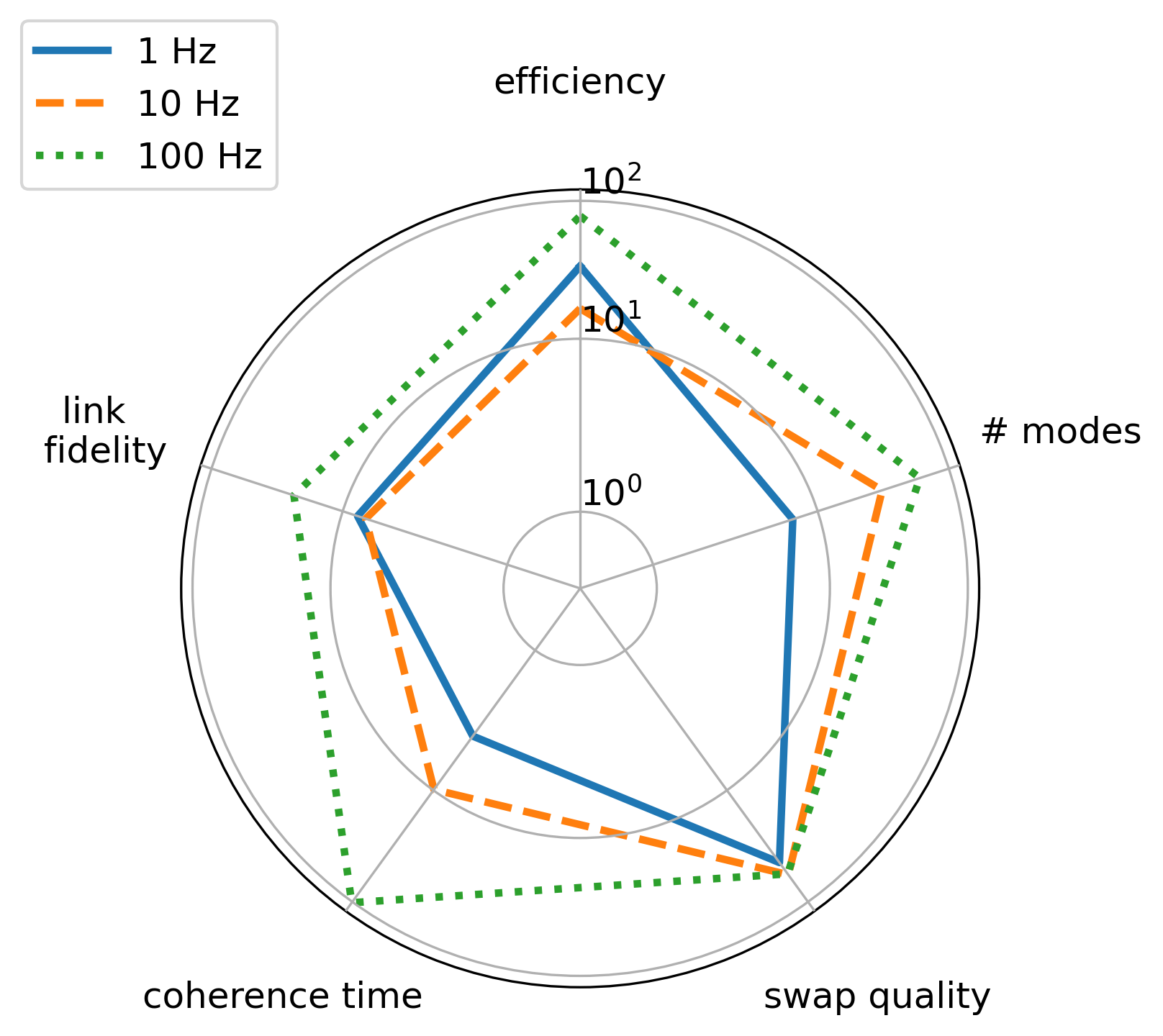}
\caption{Directions along which hardware must be improved to enable attaining secret-key rates of 1 (blue, full), 10 (orange, dashed) and 100 Hz (green, dotted) between the German cities of Bonn and Berlin.
The further away the line is from the center of the plot towards a given parameter, the more that parameter must be improved with respect to the current state-of-the-art.
Improvement is depicted for the following parameters, clockwise from the top: overall photon detection probability excluding attenuation in fiber, number of multiplexing modes, fidelity of entanglement swap, coherence time of memory qubits and fidelity of elementary links.
Note the use of a logarithmic scale.}
\label{fig:minimal_requirements_different_key_targets}
\end{figure}

\begin{figure}[!ht]
\includegraphics[width=\columnwidth]{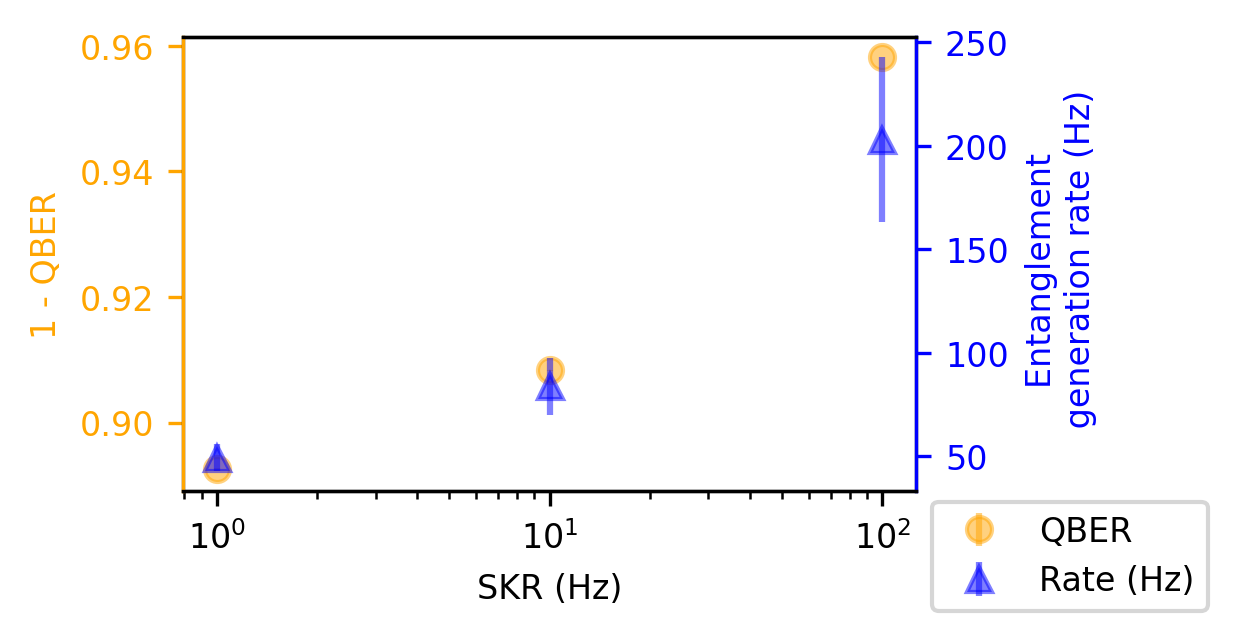}
\caption{QBER and entanglement generation rate obtained with the minimal hardware requirements to achieve 1, 10 and 100 Hz of SKR in the Bonn - Berlin setup using the configuration found to be optimal for 10 Hz.
The error bars are given by the standard error of the mean.
Each data point corresponds to 2000 simulations of an entanglement-based BB84 protocol.
}
\label{fig:qber_rate_different_targets}
\end{figure}

In Figure~\ref{fig:minimal_requirements_different_key_targets} we show the directions along which hardware must be improved for distributing secret key at rates of 1, 10 and 100 Hz using BB84 in the network path connecting Bonn and Berlin.
The corresponding minimal hardware requirements can be found in Table~\ref{tab:minimal_requirements_values}.
The hardware requirements become more stringent as the SKR target grows.
Further, the coherence time requires significantly less improvement in the 1 Hz case when compared to 10 and 100 Hz.
This comes as something of a surprise, given that we expect qubits to spend less time in memory for higher SKR values, as these should correspond to higher entanglement-generation rates (and hence lower waiting times).
In order to further investigate why this happens, we show in Figure~\ref{fig:qber_rate_different_targets} the QBER and entanglement generation rate achieved with the minimal hardware requirements for the best setup found by our optimization procedure for different SKR targets.
We find that both the entanglement generation rate and $1 - \text{QBER}$ increase with the target SKR.
We conjecture that the increase in coherence time observed for higher SKR targets is due to the necessary entanglement generation rate being very high.
In fact, it is so high that it requires a huge number of multiplexing modes, which in turn imply a very high cost.
This makes it comparatively less costly to extract more key from each entangled state than to generate states faster.
\subsection{Requirements for secret-key and blind-quantum-computing success rates}
\begin{figure}[!ht]
\includegraphics[width=\columnwidth]{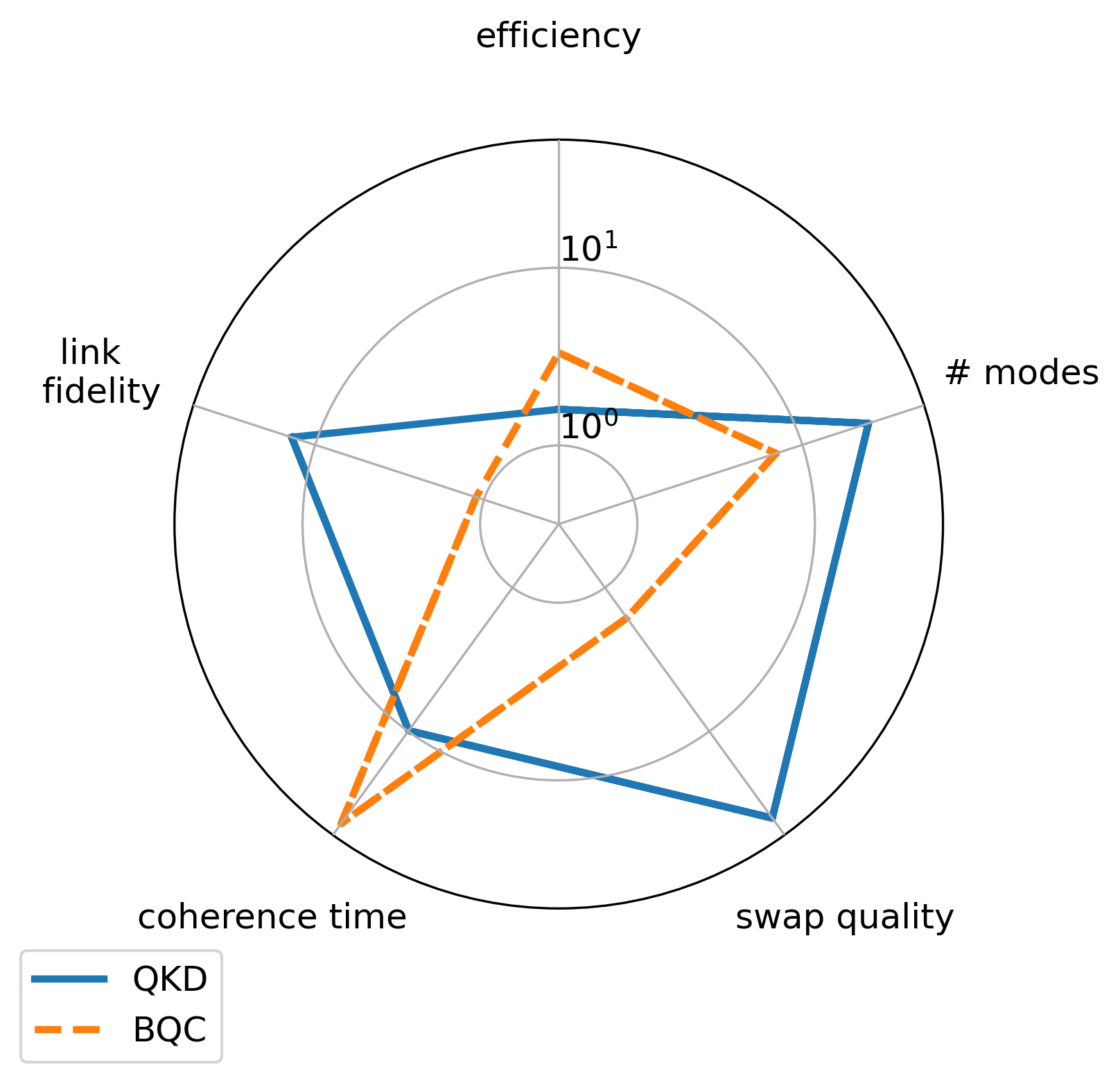}
\caption{Directions along which hardware must be improved to enable attaining secret-key (QKD, blue) and blind quantum computing (BQC, orange) test protocol rates of 10 Hz between the German cities of Bonn and Berlin.
The further away the line is from the center of the plot towards a given parameter, the more that parameter must be improved with respect to the current state-of-the-art.
Improvement is depicted for the following parameters, clockwise from the top: overall photon detection probability excluding attenuation in fiber, number of multiplexing modes, fidelity of entanglement swap, coherence time of memory qubits and fidelity of elementary links.
Note the use of a logarithmic scale.}
\label{fig:minimal_requirements_different_target_applications}
\end{figure}

In Figure~\ref{fig:minimal_requirements_different_target_applications} we show the directions along which hardware must be improved for performing QKD and BQC at a rate of 10 Hz in the network path connecting Bonn and Berlin.
The corresponding minimal hardware requirements can be found in Table~\ref{tab:minimal_requirements_values}.
It is plain to see that the two applications require improvements in distinct parameters.
In particular, we emphasize the much larger coherence time required for BQC, corresponding to roughly a factor of 2.5 difference (7.99 vs 3.14 seconds).
This can be explained by the fact that BQC, unlike QKD, requires two entangled pairs to be alive at the same time, implying that one entangled pair must be stored at the end nodes while the second one is generated.
Further, the fact that the minimal coherence time required for BQC is high means that comparatively less noise will be caused by decoherence.
This, in turn, means that in order to achieve the same state quality, the swap quality and the elementary link fidelity need not be as good.

We have also observed that there is a significant difference in the entanglement generation rates achieved by the parameter sets corresponding to the improvements shown in Figure~\ref{fig:minimal_requirements_different_target_applications}.
The minimal hardware requirements for QKD achieve an entanglement generation rate of almost 80 Hz, whereas the ones for the BQC-test-protocol result in an entanglement generation rate of around 20 Hz.
In the same vein as what was discussed in Section~\ref{sec:qber_rate_tradeoff}, this is a result of the SKR and the BQC test protocol success rate being composite quantities, depending not only on the rate at which entangled states are delivered, but also on the quality of these states.
We believe that the difference observed in entanglement generation rate between the two applications is due to the fact that there is a threshold state quality to obtain non-zero secret-key ($\sim 11\%$ QBER or equivalently $\sim 0.84$ fidelity, both under the assumption of depolarizing noise).
Such a threshold does not exist for the BQC test protocol.
This fundamental difference means that the state quality requirements are more stringent in the QKD case, making improving the entanglement generation rate a more attractive possibility.
We do however note that even though the BQC test protocol does not impose a threshold on state quality, the complete VBQC protocol proposed in~\cite{leichtle2021verifying} does.

\section{Conclusion}\label{sec:conclusion}
We have determined minimal hardware requirements for generating entanglement between two nodes separated by roughly 900 km of real-world optical fiber using a chain of processing-node quantum repeaters.
We investigated both how such requirements depend on how many repeaters are employed and on the quantum-network application for which the entanglement is used.
Notably, we have found that the hardware requirements for performing quantum key distribution and a simplified form of blind quantum computing are qualitatively different, with blind quantum computing requiring a coherence time which is roughly a factor of 2.5 larger for the same target rate in the setup we investigated.
We further observed that given that most metrics one is interested in when evaluating quantum-network performance depend on both the rate at which entanglement is generated and its quality, there is room for trade-offs:
for example, we found that when employing a large number of repeaters to achieve a given secret-key rate in the setup we studied it is easier to generate many entangled pairs of comparatively lower quality, with the opposite being true if fewer repeaters are used.

The blind-quantum-computing requirements we determined were obtained for a simplified form of the protocol, which is useful as a benchmark for quantum-network performance but is not an interesting application in and of itself.
It would be interesting to learn how the results presented would change if instead a complete verified blind quantum computing protocol such as the one introduced in~\cite{leichtle2021verifying} were studied.


\section{Data availability} \label{sec:data_availability}
The data presented in this work have been made available at \url{https://doi.org/10.4121/22193539}~\cite{2023requirementsreplicationdata}.

\section{Code availability} \label{sec:code_availability}
The code that was used to perform the simulations and generate the plots in this paper has been made available at \url{https://gitlab.com/softwarequtech/simulation-code-for-requirements-for-upgrading-trusted-nodes-to-a-repeater-chain-over-900-km-of-optical-fiber}~\cite{deutsche_telekom_code}.

\section*{Acknowledgements} \label{sec:acknowledgements}
We thank Bethany Davies for critical reading of the manuscript.
We thank Deutsche Telekom for sharing data regarding their fiber network.
This work was supported by the QIA-Phase 1 project which has received funding from the European Union’s Horizon Europe research and innovation programme under grant agreement No. 101102140.
G.A. was supported by NWO Zwaartekracht QSC 024.003.037.

\bibliography{biblio} 
\bibliographystyle{ieeetr}
\onecolumngrid
\appendix*
\appendix
\onecolumngrid
\section{Baseline parameters}
\label{sec:appendix_baseline_parameters}
Here we discuss how we determined the baseline hardware parameters shown in Table~\ref{tab:baseline_parameters}.
We did so by following two steps:
(i) finding state-of-the-art color-center hardware parameters in the literature and 
(ii) converting these to the hardware model we employ.
In Table~\ref{tab:color_center_parameters} we show the relevant state-of-the-art color center parameters we have identified and provide their respective references.
\begin{table}[!htpb]
\begin{tabular}{|c|c|}
\hline
Parameter                               & State-of-the-art value \\ \hline
Number of modes                         & 1                      \\ \hline
Carbon coherence time                          & 1 s~\cite{bradley2019ten} \\ \hline
Elementary-link fidelity                & 0.83~\cite{hensen2015loophole}                   \\ \hline
Electron initialization fidelity        & 0.995~\cite{pompili2021realization}                  \\ \hline
Carbon initialization fidelity          & 0.99~\cite{bradley2019ten}                   \\ \hline
Electron-carbon two-qubit gate fidelity & 0.97~\cite{kalb2017entanglement}                   \\ \hline
Electron single-qubit gate fidelity     & 0.995~\cite{pompili2021realization}                  \\ \hline
Carbon single-qubit gate fidelity       & 0.999~\cite{taminiau2014universal}                  \\ \hline
Electron readout fidelity               & 0.93(0) 0.995(1)~\cite{hermans2022qubit}       \\ \hline
Photonic interface efficiency           & 0.855~\cite{bhaskar2020experimental}                   \\ \hline
Frequency conversion efficiency         & 0.3~\cite{zaske2012visible}                    \\ \hline
\end{tabular}
\caption{State-of-the-art color center parameters.
We have number of modes as $1$ without reference because to the best of our knowledge multiplexed entanglement generation has not been demonstrated using color centers.}
\label{tab:color_center_parameters}
\end{table}
We now discuss how these are converted to the parameters shown in Table~\ref{tab:baseline_parameters}.
The elementary-link fidelity and number of modes can be used directly without conversion.
Color-center memories have both an electron qubit (also known as communication qubits due to their optical interface) and possibly multiple carbon qubits (also known as memory qubits due to being long-lived).
We assume a `best-of-both-worlds' situation, in which the qubits in our model are both endowed with an optical interface that allows them to generate entanglement and a long ($1$s baseline) memory lifetime.
This simplification allows us to treat all qubits in the nodes equally.
As explained in Section~\ref{sec:setup} we combine all photon-related inefficiencies, with the exception of fiber attenuation, into one parameter, $p_{\text{det}}$.
This is done as follows:
\begin{equation}
p_{\text{det}} = p_{\text{photon interface}} \cdot p_{\text{conv}}, 
\end{equation}
where $p_{\text{photon interface}}$ is the photonic interface efficiency and $p_{\text{conv}}$ is the frequency-conversion efficiency.
This results in the $0.255$ number reported in Table~\ref{tab:baseline_parameters}.
We note that the experiment reported in~\cite{bhaskar2020experimental} does not consist of entanglement generation through a heralding station, as we assume in this paper.
We have made a best guess of how the parameters reported there would translate to a scheme where entangled photons are interfered and measured at a heralding station.
An entanglement swap in a color center (this concrete example was demonstrated using a nitrogen-vacancy center) consists of single-qubit gates on both carbon and electron, two-qubit gates and measurement and initialization of the electron (see Figure 17 in Supplementary Note 5 of~\cite{coopmans2021netsquid} for an image of the circuit).
We make the simplifying assumption that all errors are depolarizing.
First, we convert each of the initialization and gate fidelities in Table \ref{tab:color_center_parameters} to depolarizing parameters (in accordance with Equation \eqref{eq:depolarizing}),
and then multiply the depolarizing parameters corresponding to all the operations in the circuit together to obtain the swap quality (which parametrizes a depolarizing channel as detailed in Section~\ref{sec:setup}), i.e.,
\begin{equation}
    s_q = (1 - p_{\text{carbon}})^2 \cdot (1 - p_{\text{electron-carbon}}) \cdot (1 - p_{\text{electron}})^2 \cdot (1-p_{\text{electron init}}) \cdot (1-p_{\text{electron meas}})^2 \cdot (1-p_{\text{retrieve}}),
\end{equation}
where $p_{\text{carbon}}$ is the depolarizing parameter of the carbon single-qubit gate, $p_{\text{electron-carbon}}$ of the two-qubit gate, $p_{\text{electron}}$ of the electron single-qubit gate, $p_{\text{electron init}}$ of the electron initialization, $p_{\text{electron meas}}$ of the electron measurement and $p_{\text{retrieve}}$ of the retrieve operation (maps the carbon state to the electron, see Figure 17 (b) in Supplementary Note 5 of~\cite{coopmans2021netsquid}).

\section{Repeater placement chosen by optimization method}
\label{sec:appendix_best_rep_placement}
\begin{figure}[!ht]
\centering
\includegraphics[width=0.75\columnwidth]{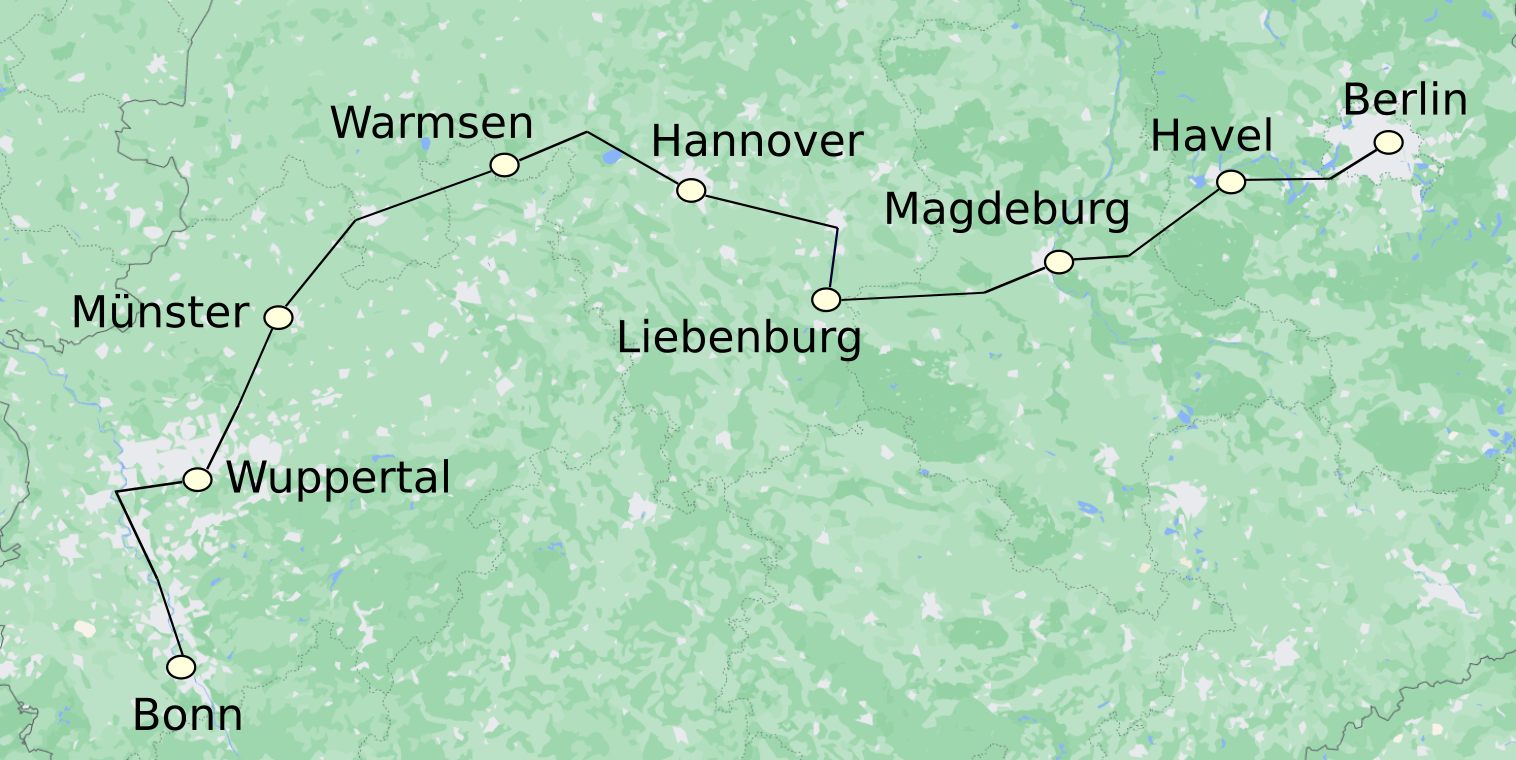}
\caption{Map of Germany overlaid with a depiction of the fiber path connecting the German cities of Bonn and Berlin that we investigated.
The white circles represent end nodes, in Bonn and Berlin, and repeater nodes elsewhere.
This placement corresponds to the best found by our optimization method, in the sense that it allowed for minimization of hardware requirements for a target secret-key rate of 10 Hz.
}
\label{fig:bonn_berlin_best_config}
\end{figure}

As described in Section~\ref{sec:min_requirements_qkd}, we determined minimal hardware requirements for performing QKD at a rate of 10 Hz over the network path depicted in Figure~\ref{fig:bonn_berlin_satellite}.
In doing so, we optimized over the number of repeaters used and their placement.
We then used the placement our optimization method found to perform best for determining minimal hardware requirements for other performance targets, as described in Section~\ref{sec:results_target}.
In Figure~\ref{fig:bonn_berlin_best_config} we show this placement.
In Table~\ref{tab:details_best_path} we show the lengths and attenuations of the elementary links defined by the repeater placement.
\begin{table}[!ht]
\begin{tabular}{|c|c|c|}
\hline
Link                   & Length (km) & Attenuation (dB) \\ \hline
Bonn - Wuppertal       & 138.9       & 32.8             \\ \hline
Wuppertal - Münster    & 133.2       & 31.4             \\ \hline
Münster - Warmsen      & 126.2       & 29.6             \\ \hline
Warmsen - Hannover     & 97.2        & 22.7             \\ \hline
Hannover - Liebenburg  & 122.0       & 28.4             \\ \hline
Liebenburg - Magdeburg & 115.5       & 26.9             \\ \hline
Magdeburg - Havel      & 103.9       & 24.3             \\ \hline
Havel - Berlin         & 80.2        & 18.6             \\ \hline
\end{tabular}
\caption{Length and attenuation of elementary links depicted in Figure~\ref{fig:bonn_berlin_best_config}.}
\label{tab:details_best_path}
\end{table}

In order to optimize over the number of repeaters and their placement, we have first generated all the 986 possible ways repeaters can be placed in the network (such that there is still space for the required heralding stations between repeaters and end nodes).
To each configuration we assigned a number $r$ corresponding to the number of repeaters in the network.
Then, for each configuration we computed the chain asymmetry parameter defined as \cite{avis2023asymmetry}
\begin{equation}
\mathcal A_\text{chain} = \frac 1 r \sum_{i=1}^r \frac{|L_{\text{left}, i} - L_{\text{right}, i}|}{L_{\text{left}, i} + L_{\text{right}, i}},
\end{equation}
where $L_{\text{left}, i}$ ($L_{\text{right}, i}$) is the distance between repeater node $i$ and its left- (right-) hand neighboring node.
Next, we ordered all the configurations with the same value of $r$ by their values of $\mathcal A_\text{chain}$, and label their position in this ordering as $n$.
This number is then an identifier for how asymmetric (as quantified by the chain asymmetry parameter) a configuration is relative to the other configurations with the same number of repeaters.
$n=0$ corresponds to the most symmetric setup and $n=m_r - 1$ corresponds to the most asymmetric setup, where $m_r$ is the number of configurations with $r$ repeaters.
All configurations are then stored in a table by their values for $r$ and $n$.

Then, when we optimize over the hardware parameters, we also optimize over two additional parameters.
These are $r$ (the number of repeaters) and $a$, which is a number between zero and one.
Given a pair $(r, a)$, the configuration that is used is chosen as follows.
First, the number $a$ is mapped to a value of $n$ using
\begin{equation}
n = \text{round}\Big(a (m_r - 1)\Big),
\end{equation}
(where round denotes rounding to the closest integer)
i.e., it uses $n=0$ for $a=0$ and $n= m_r -1$ for $a = 1$.
Second, the unique configuration defined by the values of $r$ and $n$ is taken from the table and used in the simulation.
The reason why we optimize over $a$ instead of over $n$ directly is that $a$ quantifies how asymmetric the chosen configuration is in a way that is independent of $r$ (the range is always between 0 and 1, instead of between 0 and $m_r - 1$).
This makes it easier to vary $r$ and $a$ independently compared to $r$ and $n$.

\section{Optimization method}
\label{sec:optimization_method}
In this appendix we provide more details regarding our optimization methodology.
This methodology is based on genetic algorithms, which come in several different flavors.
Our particular implementation is heavily based on the one introduced in~\cite{da2021optimizing} and used in~\cite{avis2022requirements, labay2021genetic, labay2023genetic} to which we refer the interested reader.
There are two things that we do discuss in this appendix.
First, as the parameter set we use here is different than in \cite{da2021optimizing, avis2022requirements, labay2021genetic, labay2023genetic}, we explain in Section \ref{app:no_imperfection_probs} how we define the probability of no imperfection for each of these, as required by the definition of the hardware cost function $H_c$ in Section \ref{sec:defining_minimal_hardware_requirements}.
Second, we have employed a simple local optimization performed on the best solution found by the genetic algorithm.
This local optimization method has been used and described in~\cite{labay2021genetic}, but has not yet appeared in a peer-reviewed publication.
We therefore explain it below in Section \ref{app:local_optimization}.
Additionally, we also give the details of the machine used to perform the actual optimizations in Section \ref{app:performing_optimization}.
Finally we would like to remark that the code for our implementation, together with the tools required for integration with NetSquid simulations, is publicly accessible at~\cite{smartstopos}.

\subsection{No-imperfection probabilities}
\label{app:no_imperfection_probs}
We show in Table~\ref{tab:probabilities_no_error} the probability of no-imperfection for all parameters considered in our hardware models.
\begin{table}[!ht]
\begin{tabular}{|c|c|}
\hline
Parameter                                                             & Probability of no-imperfection                      \\ \hline
Photon detection probability $p_{\text{det}}$                         & $p_{\text{det}}$                             \\ \hline
Coherence time $T$                                              & $e^{- 1/T}$                              \\ \hline
Swap quality $s_q$                                                    & $s_q$                                        \\ \hline
Elementary link fidelity $F_{el}$                                                    & $F_{el}$                                        \\ \hline
Number of multiplexing modes $N$                                                    & $1 - \left(1 - p_{\text{surv baseline}}\right)^N$                                        \\ \hline
\end{tabular}
\caption{Probabilities of no-imperfection for hardware parameters we optimized over in this work.
$p_{\text{surv baseline}}$ (defined in Equation \eqref{eq:p_surv_baseline}) is the probability of one photon (i.e., no multiplexing) surviving traveling an elementary link made up out of two times the average fiber segment in the fiber path we study (shown in Figure~\ref{fig:bonn_berlin_satellite}).
}
\label{tab:probabilities_no_error}
\end{table}

We start by defining the quantity $p_{\text{surv baseline}}$ that appears in this table more rigorously.
It is computed as follows,
\begin{equation} \label{eq:p_surv_baseline}
    p_{\text{surv baseline}} = 10^{-\overline{\alpha_{\text{att}}}/10},
\end{equation}
with $\overline{\alpha_{\text{att}}}$ given by,
\begin{equation}
    \overline{\alpha_{\text{att}}} = \frac 2 N  \sum_{i=1}^N \alpha_{\text{att}, i} L_i.
\end{equation} 
Here, $L_i$ is the length of fiber segment $i$ in the fiber path under consideration, $\alpha_{\text{att}, i}$ is the attenuation coefficient of fiber segment $i$ (i.e., the amount of attenuation per unit length), and $N$ is the total number of fiber segments in the path.
For the fiber path considered in this paper (i.e., the one depicted in Figure \ref{fig:bonn_berlin_satellite}), $N = 17$.
An elementary link between two neighboring nodes must consist of at least two fiber segments to allow for the installation of a heralding station.
$\overline{\alpha_{\text{att}}}$ can then be thought of as the total amount of attenuation on an elementary link made up of two times the average fiber segment.
This means $p_{\text{surv baseline}}$ is the probability of a photon surviving traveling through this average elementary link.
The reason for constructing this quantity is that it provides a baseline for the photon survival probability in fiber, which can then be improved upon by increasing the number of multiplexing modes, thereby enabling us to associate a cost function.

The coherence time $T$ represents a timescale for depolarization, with the probability of the state becoming maximally mixed over a period of time $t$ being given by $1 - e^{-t/T}$, with the respective probability of no-imperfection then being $e^{-t/T}$.
In this case, improving $T$ by a factor of $k$ is equivalent to multiplying it by $k$.

For the swap quality, $s_q$ is the probability that the two-qubit state before the Bell-state measurement is not replaced with a maximally mixed state, and therefore $s_q$ is the corresponding probability of no imperfection.
Finally, for the elementary-link fidelity we take the fidelity itself to be the probability of no imperfection.

\subsection{Local optimization}
\label{app:local_optimization}
Genetic algorithms are derivative-free optimization algorithms that are particularly useful when applied to functions whose cost landscape is largely unknown but is assumed to have many local minima~\cite{10.5555/534133}.
Through a balancing act of exploration (i.e., investigation of many different areas of parameter space) and exploitation (i.e., investigation of local optima) they often manage to avoid being trapped in local optima as gradient-based methods are wont to.
Nevertheless, use of a genetic algorithm does not guarantee that one can find the global optimum.
Further, one can not even be sure that one has maximally exploited the best optimum found.
For this reason, we complement the exploration performed by the genetic algorithm with a deterministic local optimization method which we apply to the best parameter set found by the genetic algorithm.
The algorithm used is a variation of an iterative local search algorithm~\cite{luke2013essentials}.
It consists of iteratively making small changes on a parameter and evaluating the cost associated to the resulting parameter set.
In case it has decreased, it is kept and we again make a small change on the same parameter.
If the cost increases, we discard the change and move on to another parameter.
This process is repeated for all parameters being optimized over.
We must however emphasize that this still does not guarantee that the global optimum will be found.

More details on this method can be found in Chapter 4.2 of~\cite{labay2021genetic}. 

\subsection{Performing the optimization}
\label{app:performing_optimization}
Each optimization run was executed on a thin node of the Snellius supercomputer~\cite{snellius}.
Each of these nodes is endowed with 2 AMD Rome 7H12 CPUs (2.6 GHz), for a total of 128 cores and a total of 256 GiB of memory.

\section{BQC test protocol}
\label{app:bqc_test_protocol}

In this appendix we describe the BQC test protocol that is used as a performance metric in this paper.
This protocol consists of repeated execution of test rounds as required by the VBQC protocol presented in~\cite{leichtle2021verifying}.
In each round of the VBQC protocol, a server is tasked by a client to execute a quantum computation on qubits transmitted by the client and then send the classical result of that computation back to the client.
In test rounds, the client has prepared the transmitted qubits in such a way that it knows the correct outcome of the computation.

Therefore, executing test rounds allows the client to verify whether the server is honest.
However, noise in the quantum hardware can also lead to failed test rounds.
The more often test rounds fail due to noise, the harder it is for the client to verify the server's honesty.

The BQC test protocol that we consider is not itself a VBQC protocol.
In fact, its only purpose is to benchmark how suited a quantum network could be to perform BQC protocols (and perhaps other applications that require multiple live qubits simultaneously).
The performance metric that we consider for this protocol is the success rate, defined as the average number of successful test rounds that can be executed per time unit (i.e., the product of the rate $R$ and success probability $p_s$, as in Equation \eqref{eq:BQC_success_rate}).
We specifically consider an entanglement-based two-qubit version of the protocol.
In that case, the protocol is as follows:
\begin{enumerate}
\item
The client chooses $d$ and $r$ uniformly at random from $\{0, 1\}$ and $\theta$ from $\{j\pi/4\}_{0 \leq j \leq 7}$, and then defines two quantum states,
$\ket{\text{dummy}} = \ket i$ and $\ket{\text{trap}} = \ket{+_\theta}$, where $\ket{\pm_\phi} \equiv \tfrac 1 {\sqrt 2} (\ket 0 \pm e^{i\phi} \ket 1)$.
Additionally, it uniformly at random designates $\ket{\phi_1}$ to be $\ket{\text{dummy}}$ or $\ket{\text{trap}}$.
$\ket{\phi_2}$ is designated to be the option that was not chosen.

\item
When an entangled state shared between the client and server becomes available, the client uses quantum teleportation to transmit the state $\ket{\phi_1}$ to the server.
The server stores the received state in quantum memory.

\item
When a second entangled state becomes available, the client uses quantum teleportation to transmit the state $\ket{\phi_2}$ to the server.

\item
The server performs a CZ gate between its two qubits.

\item
The server measures the qubit that was used to receive the state $\ket{\text{trap}}$ in the basis $\{\ket{+_{\theta + r \pi}}, \ket{-_{\theta + r \pi}} \}$ and transmits the result back to the client.

\item
The client declares the test round a success if it a receives measurement result matching its expectation (i.e., if the outcome is equal to $d \oplus r$, where $\oplus$ is addition modulo two), and a failure otherwise.

\item
The client and server go back to Step 1 to start the next test round.
\end{enumerate}
Alternatively, remote state preparation~\cite{bennett2001remote} could be used to prepare the required states at the server, which may be easier to execute on real hardware than quantum teleportation.
In fact, it is proven in~\cite{avis2022requirements} that using remote state preparation for the VBQC protocol in~\cite{leichtle2021verifying} is equivalent to using quantum teleportation in case the client and server implement gates noiselessly.
Therefore the success rate will, under these assumptions, be the same whether quantum teleportation or remote state preparation is used.

We here assume that classical communication between the client and the server happens instantaneously and that both the client and server are able to perform gates and measurements noiselessly and instantly.
However we do not assume they are able to store qubits indefinitely; the first teleported state undergoes depolarizing noise as described in Equation \eqref{eq:memory_decoherence}, where the coherence time $T$ is the same as the coherence time of the repeater nodes (i.e., it is varied by the optimizations performed in this paper).
Under these assumptions, $R_{rounds}$ is simply half the rate at which entanglement can be distributed when entanglement is being generated continuously, as one test round can be performed for every two entangled states that are produced.
In order to calculate the success probability, we use the following result from~\cite{avis2022requirements}:
\begin{equation} \label{eq:BQC_success_probability}
1 - p_s = e^{- \frac{\Delta t}{T}} \left[ F_\text{dummy} (1 - F_\text{trap}) + F_\text{trap} (1 - F_\text{dummy}) \right] + \frac 1 2 (1 - e^{- \frac{\Delta t}{T}}).
\end{equation}
Here, $\Delta t$ is the time between the transmission of the first qubit and the second qubit.
For the fidelities $F_\text{trap}$ and $F_\text{dummy}$,
let the density matrices for the state $\ket{\text{dummy}}$ after transmission to the server be $\rho_\text{dummy}$ and $\rho_\text{trap}$ for $\ket{\text{trap}}$.
Then $F_\text{dummy} = \bra{\text{dummy}}\rho_\text{dummy} \ket{\text{dummy}}$ and $F_\text{trap} = \bra{\text{trap}}\rho_\text{trap} \ket{\text{trap}}$.

We then determine the success rate as follows.
First, we simulate continuous entanglement generation between the end nodes of a repeater chain.
Each time an end-to-end entangled state is generated it is removed from the simulation and stored as raw data, together with the time at which it was generated.
Then, after the simulation has finished, we process the raw data to determine what the success rate would have been if the entangled states had been consumed by the BQC test protocol.
To this end, we divide the data into single test rounds, each consisting of two entangled states that were generated in succession.
We assign each test round a duration $t$, which is the amount of time between the start of the round and the end of the round (i.e., when the second state was generated), and a storage time $\Delta t$, which is the time between when the first entangled state and the second entangled state were generated.
We furthermore calculate the $p_s$ of that round using Equation \eqref{eq:BQC_success_probability}, where we average over the two possible choices in the protocol for how $\ket{\phi_1}$ and $\ket{\phi_2}$ are designated (i.e., whether the first entangled state is used to transmit the dummy and the second to transmit the trap or vice versa).
Then we calculate the rate as
\begin{equation}
R = \frac 1 {\expectationvalue{t}},
\end{equation}
where $\expectationvalue{t}$ is the average value of $t$ over all the test rounds.
Finally, we use $R$ and the average value of $p_s$ to calculate the success rate according to Equation \eqref{eq:BQC_success_rate}.
The processing code that realizes this calculation has been made publicly available at~\cite{netsquid-simulationtools}.

\end{document}